\documentclass[a4paper]{aa}

\usepackage{txfonts}
\usepackage{graphicx}
\usepackage[british]{babel}

\newcommand{\etal}{\emph{et~al.}}
\newcommand{\eg}{\emph{e.~g.}}
\newcommand{\ie}{\emph{i.~e.}}

\newcommand{\Planck}{\textsc{Planck}}
\newcommand{\boom}{BOOMERanG}

\begin{document}

\title{ROMA: a map-making algorithm for polarised CMB data sets} 

\subtitle{} \author{Giancarlo de Gasperis\inst{1}
\and Amedeo Balbi\inst{1,2} \and Paolo Cabella\inst{1} 
\and Paolo Natoli\inst{1,2} \and Nicola Vittorio\inst{1,2}}

\institute{Dipartimento di Fisica, Universit\`a di Roma ``Tor
  Vergata'', via della Ricerca Scientifica 1, I-00133, Roma, Italy
  \and Sezione INFN Roma 2, via della Ricerca Scientifica 1, I-00133,
  Roma, Italy}

\offprints{giancarlo.degasperis@roma2.infn.it}

\date{Received / Accepted}

\abstract{We present ROMA, a parallel code to produce joint optimal
  temperature and polarisation maps out of multidetector CMB
  observations. ROMA is a fast, accurate and robust implementation of
  the iterative generalised least squares approach to map-making. We
  benchmark ROMA on realistic simulated data from the last,
  polarisation sensitive, flight of \boom.}

\authorrunning{G. de Gasperis \etal}
\titlerunning{ROMA: a map-making algorithm for polarised data sets}

\maketitle 

\section{Introduction}\label{intro}
A direct consequence of the presence of anisotropies in the Cosmic
Microwave Background (CMB) radiation is a certain degree of
polarisation (\cite{1968ApJ...153..L1}, \cite{1983MNRAS...}).
Constraining CMB polarisation provides valuable cosmological
information, complementary to that encoded in temperature anisotropy
and it significantly tightens the constraints on cosmological
parameters (\cite{1996ApJ...469..437S}).

Unfortunately, the polarised component of the CMB is expected to be
small compared to total intensity, making its measurement an
experimental challenge. For this reason, until very recently the
experimental effort has not focused on the polarised component.
However, the situation is quickly changing. A first detection of CMB
polarisation, in agreement with theoretical predictions, has been
announced by the interferometric experiment DASI
(\cite{2002Natur.420..772K}).  The WMAP satellite
(\cite{2003ApJS..148....1B}) has measured the predicted correlation
spectrum between CMB temperature anisotropy and polarisation
(\cite{2003ApJS..148....1B}). Many other CMB polarisation experiments
(DASI, CBI, CAPMAP, BICEP, QUEST, \boom\ 2K, MAXIPOL, SPORT, BarSPORT,
\Planck, etc.\footnote{See \eg\ the site \texttt{lambda.gsfc.nasa.gov}
  for up to date references to ongoing and forthcoming Polarisation
  experiments}) have either already taken data or are expected to do
so in the very near future.

It is well known that in CMB science the data reduction process
requires almost as much care as data gathering.  The analysis of
temperature data has now reached full maturity, and it has been proven
successful on several datasets. The same is not true for polarisation.
Although the overall analysis scheme can be borrowed from the
temperature case, and many of the algorithms involved admit a more or
less straightforward generalisation, polarised data sets present
peculiar aspects that call for specific treatments. One basic problem
is that the tensor-like nature of polarisation poses some constraints
on the measurement process (see \cite{Cabella} or \cite{Wandelt} for a
recent review). As a consequence, not all possible instrumental
configurations are equally advantageous from the point of view of data
reduction (\cite{1999A&AS..135..579C}). The so called ``optimal''
configurations comprise several different polarisation sensitive
detectors (polarimeters), placed at a convenient reciprocal
orientation. Usually data streams from different polarimeters are
jointly analysed to seek for a faint signal. The difficulties involved
in this process, and their relation to the design of robust and
efficient data mining algorithms, are seldom considered in the
literature (see~\cite{2000A&AS..142..499R} for a noticeable
exception).

Two major steps in the data analysis process are (1) the production of
sky maps from a set of Time Ordered Data (TOD) and (2) the extraction
of the angular power spectrum from such maps. Here we focus on the
first step, leaving the discussion about power spectrum estimation to
a forthcoming paper. In a previous paper (\cite{natoli}) we described
the implementation of an optimal map-making algorithm for the case of
temperature only TOD collected by a one horned CMB experiment, and we
showed the feasibility of the method by applying it to \Planck\ and
\boom\ simulated data. Here, we discuss the generalisation of this
algorithm to the case of TOD produced by an arbitrary number of
polarisation sensitive detectors. The software implementation of this
method, which we call ROMA (Roma Optimal Map-making Algorithm), allows
fast and efficient production of optimal multidetector maps of CMB
total intensity and polarisation. The code is currently being used to
analyse the polarised data set from the second Antarctic flight of the
\boom\ experiment (hereafter B2K, \cite{montroy}), which took place in
January 2003.  ROMA is part of a full, end-to-end data analysis
pipeline for B2K, completely developed in Rome, whose details will be
described elsewhere.

The plan of this paper is as follows: In Sect.~2 we derive the least
squares equations for multidetector map-making; in Sect~3 we describe
the details of the implementation, while in Sect.~4 we report the
application of ROMA to highly realistic B2K simulated data. Finally,
in Sect.~5, we draw our main conclusions.

\section{Polarised Data Map-Making}\label{pol_data_mm}

In this section we derive the Generalised Least Squares (GLS)
equations for the polarised map-making problem, given an arbitrary
number of polarimeters. With ``polarimeter'' we mean, here and in the
following, a generic detector measuring total intensity plus a linear
polarisation component. Other types of detectors do exist and rely on
different strategies to measure polarisation. We prefer to focus on
the linear polarimeter case because of its widespread adoption. In any
case it would be straightforward to generalis of scheme once the data
model is properly modified.

The sky signal seen by a polarimeter can be expressed
as~(\cite{Chandrasekhar}):
\begin{equation}
  \mathcal{D}= \frac{1}{2}\left( I + Q \cos 2\xi + U \sin2\xi \right).
\label{eq:sign_measured}
\end{equation}
where $I$, $Q$ and $U$ are the Stokes parameters for total intensity
and linear polarisation, and the angle $\xi$ identifies the
polarimeter orientation with respect to the chosen celestial frame.
Note that we do not include the contribution of circular polarisation,
associated to the Stokes parameter $V$. As a consequence, a $V$ signal
would be seen by our polarimeter as a contribution to $I$. This fact
however is not a problem in CMB science because circular polarisation
is not generated by Thomson scattering of unpolarised radiation.

All three relevant Stokes parameters can be extracted by either
combining the output of many detectors with different mutual
orientation, or by allowing enough focal plane rotation. The data
stream output of a given detector is a combination of sky signal and
instrumental correlated noise:
\begin{equation}
  \mathcal D_{t} = \frac{1}{2} 
  A_{tp} \left( I_p + Q_p \cos 2\phi_{t} + 
    U_p \sin 2\phi_{t} \right) + n_{t}
\label{eq:TOD_TQU}
\end{equation}
Here the index $t$ labels time while $p$ runs over the pixelized
images of $I$, $Q$ and $U$.  The pointing matrix $A_{tp}$ couples the
time and pixel domain. The information about beam smearing can, in
general, be included in this matrix.  We prefer, however, to assume
that $A$ is pointwise (\ie\ having a single nonzero entry per row,
occurring when a pixel falls into the line of sight) and hide the beam
in the $I$, $Q$, and $U$ map (\ie\ solve for a pixelized, beam smeared
image of the sky). This is only meaningful if the beam is, to good
approximation, symmetric with respect to boresight, and common to $I$,
$Q$ and $U$ (see~\cite{2004PhRvD..70l3007A} for an algorithm that
takes into account the effect of an asymmetric beam). In
Eq.~(\ref{eq:TOD_TQU}) above $n_{t}$ is a vector of correlated noise.

By inserting the trigonometric functions within the pointing matrix,
and considering a set of $k$ polarimeters, one can recast
Eq.~(\ref{eq:TOD_TQU}) into a more compact formalism:
\begin{equation}
\mathbf {\mathcal D}_t = \mathbf A_{tp} \mathbf S_p + \mathbf n
\label{eq:TODgen}
\end{equation}
where the datastreams of each polarimeter are combined:
\begin{displaymath}
  \mathbf {\mathcal D}_t \equiv
  \left(
    \begin{array}{c}
      {\mathcal D_t}^1 \\
      \vdots \\
      {\mathcal D_t}^k
    \end{array}
  \right),
\end{displaymath}
and the generalised pointing matrix becomes:
\begin{displaymath}
  \mathbf A_{tp} \equiv
  \frac{1}{2} \left(
    \begin{array}{ccc}
      A_{tp}^1 &\cos 2 \phi_t A_{tp}^1 & \sin 2 \phi_t A_{tp}^1 \\
      \vdots  & \vdots & \vdots \\
       A_{tp}^k &\cos 2 \phi_t A_{tp}^k & \sin 2 \phi_t A_{tp}^k \\
    \end{array}
  \right).
\end{displaymath}
Similarly, the sky signal can be expressed as:
\begin{displaymath}
\mathbf S_p \equiv 
  \left(
    \begin{array}{c}
      I_p \\
      Q_p \\
      U_p
    \end{array}
    \right),
\end{displaymath}
while the noise stream is:
\begin{displaymath}
\mathbf n_t \equiv
  \left(
    \begin{array}{c}
      n_t^1 \\
      \vdots \\
      n_t^k
    \end{array}
\right).
\end{displaymath}
Eq.~(\ref{eq:TODgen}) defines a generic linear algebra problem, whose
unknown parameters (the map pixel values) can be constrained by means
of the standard GLS solution (\eg\ \cite{lupton}):
\begin{equation}\label{eq:MMsolution}
  \mathbf {\widetilde S_p} = \left( \mathbf A^t 
\mathbf{N}^{-1} \mathbf A\right)^{-1}
  \mathbf A^t \mathbf{N}^{-1} \mathbf {\mathcal D},
\end{equation}
with:
\begin{equation}
  \mathbf N \equiv \left\langle \mathbf n_t \mathbf n_{t^\prime}\right\rangle=
  \left(
    \begin{array}{ccc}
      \left\langle  n_t^1  n_{t^\prime}^1 \right\rangle & 
      \cdots &
      \left\langle  n_t^1  n_{t^\prime}^k \right\rangle \\
      \vdots & \ddots & \vdots  \\ 
      \left\langle  n_t^k  n_{t^\prime}^1 \right\rangle &
      \cdots &
      \left\langle  n_t^k  n_{t^\prime}^k \right\rangle 
    \end{array}
  \right),
\label{eq:gen_noise}
\end{equation}
and $\left\langle \cdot \right\rangle$ denotes the expectation value.
The $\mathbf N$ matrix becomes block diagonal when assuming that the
noise in different polarimeters is uncorrelated:
\begin{equation}
  \left\langle n_t^i n_{t^\prime}^j
  \right\rangle\ = \left\langle n_t^j n_{t^\prime}^i \right\rangle = 0 \;\;\; 
\left( i\neq j \right).
\end{equation}
We defer the actual implementation of Eq.~(\ref{eq:MMsolution}) to the
next section.


The treatment above is clearly idealistic. However, it is possible to
incorporate into the formalism the correction for a nominal level of
cross polarisation, one of the most common systematic effects
occurring in CMB polarimetry. Cross-polarisation is due to cross-talk
between the two orthogonal polarisation modes. That is, a polarimeter
may be sensitive, with efficiency $\chi_\mathrm {pol}$, to radiation
linearly polarised $90\degr$ off its natural polarisation mode. If we
assume that the cross-polarisation contamination is measurable (by
on-ground and/or in-flight tests) and it is constant across the
mission lifetime, the formalism expressed above can be generalised to
take the effect into account. If we introduce a calibration constant
$C$ and a cross-polarisation factor $\chi_\mathrm {pol}$, the data
model Eq.~(\ref{eq:TOD_TQU}) can be generalised as
\begin{eqnarray*}
  \mathcal D_{t}^i &=& \frac{C}{2} A_{tp}^i \left[ 
  \left( I_p + Q_p \cos 2\phi_{t}^i + U_p \sin 2\phi_{t}^i \right) \right. \\
&+& \left. \chi_\mathrm {pol}\left( I_p - Q_p \cos 2\phi_{t}^i - 
    U_p \sin 2\phi_{t}^i \right) \right]  + n_{t}^i
\end{eqnarray*}
that is
\begin{equation}
  \mathcal D_{t}^i = \frac{C_{\chi}}{2} 
  A_{tp}^i \left[I_p + \frac{1-\chi_\mathrm {pol}}{1+\chi_\mathrm {pol}}
\left( Q_p \cos 2\phi_{t}^i + U_p \sin 2\phi_{t}^i \right) \right] + n_{t}^i
\label{eq:TOD_xpol}
\end{equation}
where we have embedded the cross-polarisation factor in the new
calibration constant $C_{\chi}=C\left(1+\chi_\mathrm {pol} \right)$.
The new data model expressed in Eq.~(\ref{eq:TOD_xpol}) is then solved
by Eq.~(\ref{eq:MMsolution}) provided the generalised pointing matrix
$\mathbf A$ is rewritten accordingly. Note also that if $\chi_\mathrm
{pol}$ is close to unity , the map-making problem expressed by
Eq.~(\ref{eq:MMsolution}) becomes ill-conditioned since the matrix
$\mathbf A^t \mathbf{N}^{-1} \mathbf A$ is singular if
$\chi_\mathrm{pol}=1$. For real-life experiments we expect a
cross-polarisation level of $< 10\%$ (\eg\ \cite{montroy} for B2K),
not enough to hinder the search for a solution of
Eq.~(\ref{eq:MMsolution}). It is important to realize that the scheme
outlined above can correct for a nominal, known cross polarisation
level. Any uncertainty on this value, as well as on the overall
calibration, is a potential source of systematics.

\section{Implementation}

The straightforward implementation of Eq.~(\ref{eq:MMsolution}) for a
modern CMB experiment would require storing and inverting a huge
matrix (the map noise correlation matrix), a task often beyond the
reach of present day supercomputers. On the contrary, iterative
methods only require matrix to vector products and appear well suited
to tackle the problem. One such scheme, proposed by Natoli \etal\ 
(2001), has been shown to be particularly convenient in the case of
temperature map-making even for a high resolution full sky survey such
as \Planck. The idea (\cite{wright96}) is to decompose the product
$\mathbf A^t \mathbf{N}^{-1} \mathbf A$ (``unroll, convolve and
rebin'') to avoid computing and storing the map correlation matrix,
and make use of a preconditioned conjugate gradient (PCG) solver. The
key assumptions are: (1) assume that the beam is axisymmetric, so to
keep the structure of the pointing matrix $\mathbf{A}$ as simple as
possible and (2) assume that the noise is (piecewise, at least)
stationary and that its correlation function decays after a time lag
much shorter than the length of the timeline piece being processed.
Under the last assumption, the $\mathbf{N}$ matrix is approximately
circulant, and as such it is diagonalized by a Fourier transform. This
approach achieves linear scaling (with the number of time samples) per
PCG iteration; if the system is preconditioned by the pixel hits
counter (the number of times each pixel is seen), an accurate solution
can be obtained in a few tens of iterations (see~\cite{natoli} for a
discussion of algorithmic details; other implementations of this
  approach include \cite{dore} and \cite{madmap}).

\subsection{The IQU approach}

Our approach is to extend the above mentioned method to polarisation.
Since the timestream includes contributions from both total intensity
and linear polarisation (see Eq.~(\ref{eq:TOD_TQU})) it is desirable
to solve jointly for the temperature and polarisation maps, in the
spirit of Eq.~(\ref{eq:MMsolution}). This approach, that we call IQU,
is computationally more intensive than solving for $I$ and $(Q,U)$
separately, but preferable for at least two reasons: first, it
minimises the $Q$ and $U$ contaminations on the $I$ map; second it is
more general, because it does not require any special constraint on
the noise properties of the polarisation sensitive detectors, as
instead would be the case when the TOD are explicitly differenced to
remove the intensity component (it is easy to realize that in the last
case the matrix $\mathbf{N}$ would not be circulant except in the
special case of having identical noise properties in different
channels).

While the intensity (temperature) field can in principle be
reconstructed from a single, one horned detector (up to a level of the
same order of magnitude as the polarisation contribution to the
timeline), the same thing is not true for polarisation, unless special
care is taken in ensuring that the detector observes all pixels of the
sky under very different orientations. The latter strategy has indeed
been chosen by a few experiments (\eg\ MAXIPOL, that uses a rotating
grid, Johnson \etal\  2003), but the majority (\eg\ B2K, \Planck) plan
to extract the polarisation maps by comparing channels having
different mutual orientation of the plane sensitive to linear
polarisation. If such a strategy is used, it is necessary to reduce
several channels simultaneously in order to obtain a polarisation map,
\ie\ one is forced to perform multi-detector map-making. Of course,
multi-detector map-making can also be performed in the case of
polarisation insensitive detectors, when an optimal temperature map
has to be produced out of different channels at a given frequency (not
necessarily taken by the same experimental system).  The IQU approach
is therefore very general, as it also allows in a very natural way to
merge data taken by different channels, for temperature and
polarisation at the same time.

Due to optical constraints, different detectors do not observe exactly
the same region of the sky during the mission lifetime, with the
noticeable exception of full sky surveys. A polarisation map can only
be extracted for pixels observed with sufficient variety of focal
plane orientation, in order not to incur in an ill-conditioned
$\mathbf{A}^T\mathbf{N}^{-1}\mathbf{A}$ matrix.  Hence, it is a very
natural choice to restrict the polarisation map to intersection of the
sky regions surveyed by different detectors.  We impose this
constraint by flagging the time samples associated to pixels outside
of the common region. A more refined discrimination would entail
selecting explicitly those pixels observed with enough spread in the
polarisation angle. For the sake of simplicity we choose to restrict
the $(Q,U)$ maps to the intersection region of detector coverage. The
same constraint is neither necessary nor advantageous for the $I$ map:
it is not necessary, because the temperature is normally well defined
outside of the commonly observed region (although in general with a
larger noise contribution); it is not advantageous, because of the way
the solver works: in fact, scans would often go continuously in and
out of the common region; when unrolling a tentative solution map
$\mathbf{m}$ over the timestream (\ie\ when performing the operation
$\mathbf{Am}$), the latter would display time gaps if the solution is
restricted to the common region.  One way to tackle this problem is to
fill the gaps with a constrained noise realization.  This approach has
the disadvantage of increasing the computational cost of the
algorithm, because standard methods to produce constrained noise
realizations scale as the third power of the noise correlation length
(\cite{1991ApJ...380L...5H}). We prefer instead, to avoid the
introduction of extra gaps in the unrolled timestream, to solve for
$I$ in the whole ``union'' region, and using the $I$ tentative
solution map to obtain a continuous timestream.

\subsection{Noise estimation}
The GLS algorithm described above requires knowledge of the noise
correlation properties of each detector system. These must be
estimated from the data themselves, which consist of a combination of
noise and (often non negligible) signal. Hence, the problem of noise
estimation. One possible solution is to start from a non optimal yet
unbiased estimate of the signal (\eg\ obtained by naive coadding),
subtract it from the data and use the residual to produce an estimate
of the underlying noise power spectrum. The latter can be used to
compute a GLS map, employed in turn to produce a new noise estimate,
and the scheme can then be iterated. This is basically the approach
described in Prunet \etal\ (2000) and Dor\'e \etal\ (2001), and is
\emph{de facto} identical to a scheme proposed earlier by Ferreira \&
Jaffe (2000), although this latter algorithm is cast in a Bayesian
formalism. An alternative approach, proposed by Natoli \etal\ (2002)
is to stop the noise iteration at the first step (thus saving
computational time) and increase statistical efficiency by performing
a parametric fit on the estimated noise power spectral density.
Although computationally more efficient, this latter scheme depends
somehow on the parametric assumption for the functional form of the
noise power spectral density. For the purpose of this paper we do not
want to loose generality, and therefore we opt for estimating the
noise by means of the iterative scheme. However, the parametric
approach is a viable option for those cases where the noise spectral
density can be conveniently modelled.

In the case of multi-detector, polarisation sensitive map-making, it
is a logical choice to use the global IQU maps as an estimate of the
underlying signal. We will show below that this strategy achieves a
substantial reduction of the estimated noise residual bias, of order
the number of detector considered, as one would naively guess by
applying a heuristic argument.

%
%

A significant fraction of the data gathered from a real world
experiment are badly contaminated and cannot be used for science
extraction. Corruption of these data normally occurs because of
annoying events, either unpredictable (\eg\ cosmic ray hits) or
necessary (\eg\ the ignition of a calibrator). Data loss can also
affect pointing information: in this case, the corresponding sky
signal must be discarded even if usable \emph{per se}.  In any case,
excerption of the affected samples destroys the timeline continuity
and, hence, noise correlation. Both things must be reintroduced
somehow; this is normally done by inserting a noise realization,
constrained to reproduce the correct statistical properties while
being continuous (in a stochastic sense) at the gap boundaries (see
\cite{1991ApJ...380L...5H,2002PhRvD..65b2003S}). At the same time, the
map making code must be warned that such samples do not contain any
useful signal, a requirement trivially accomplished by associating to
all samples a timeline ``flag''. Flagged samples do not project over
the sky but are assigned to a virtual pixel, whose value is estimated
by the solver. Note that this scheme is equivalent to an ideally
modified experiment, which observes the sky and, for flagged samples,
a null signal (virtual) pixel. The latter is eventually assigned a
noise variance (and covariance to real pixels) compatible with
instrumental properties.  As shown in the next section, we find that
this approach allows for precise signal recovery in the noise free
limit, while correctly minimising the noise contribution in the weak
signal regime.

\section{Application to a realistic dataset}

To test ROMA we simulated and reduced a full mission scan (about 10
days) of the 145~GHz channels of the B2K experiment, using the nominal
pointing solution, which includes a polarisation angle for each
detector horn (de Bernardis, priv.\ comm.).

B2K (\cite{montroy}) is basically the follow up of \boom\ 
(\cite{2003ApJS..148..527C}) featuring polarisation sensitive
bolometers (PSB) and a scanning strategy optimised to measure the
polarisation of the CMB and of Galactic foregrounds. The portion of
the sky observed by B2K overlaps almost completely with the target of
the 1998 Antarctic campaign of \boom\ (\cite{2000Natur.404..955D}).
However, the scanning strategy of B2K differs from the one implemented
for the previous flight.  The survey area of B2K is split between a
$\sim 300$ square degrees region close to the Galactic plane and a low
foreground emission area of $\sim 1300$ square degrees for CMB
science. This area is in turn divided between a ``shallow'' and a
``deep'' region which differ by more than an order of magnitude in
integration time per pixel. The deep region is totally embedded in the
shallow one and covers about 120 square degrees.

\begin{figure}
  \resizebox{\hsize}{!}{\includegraphics{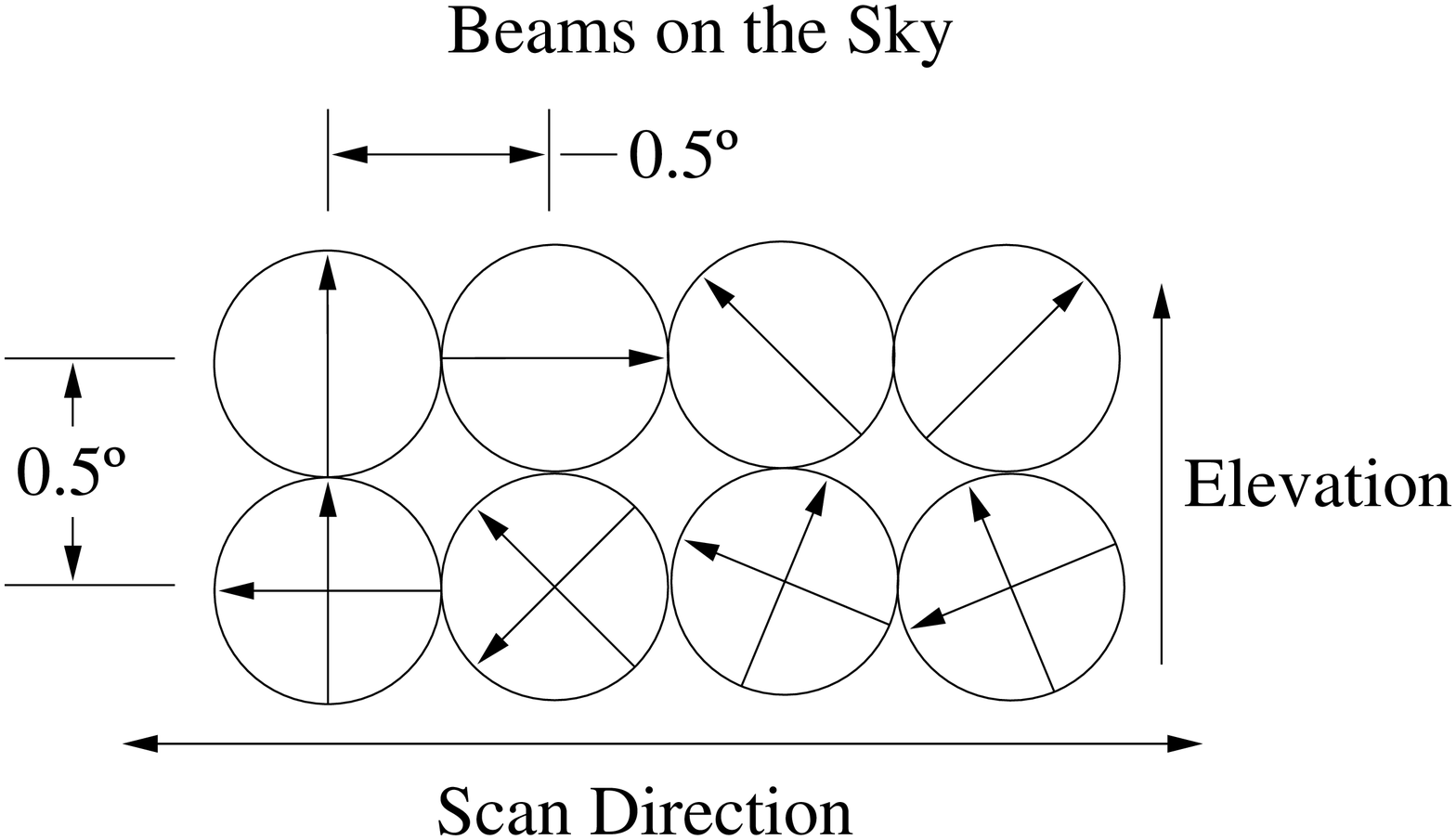}}
\caption{The B2K focal plane (\cite{masi}).
  The 145~GHz PSB's are located in the lower row. The upper detector
  row hosts unpolarised (spider web) bolometers coupled to a
  polarimeter. }
\label{fig:b2k_fp}
\end{figure}

In Fig.~(\ref{fig:b2k_fp}) we show the experimental setup of the B2K
focal plane, with the 145~GHz PSB's in the lower row.

In simulating the data the following scheme is employed: we generate a
high resolution map ($1\farcm 7$ or NSIDE=2048 in
HEALPix\footnote{\texttt{http://www.eso.org/science/healpix/}}
language) of a standard CMB sky with $\Lambda$CDM cosmology. We assume
a Gaussian beam of $10\arcmin$~FWHM for all the PSBs. The signal time
streams are then obtained using the nominal B2K scan. The noise time
streams are simulated assuming the following form for the noise
spectral density:
\begin{equation}
P(f)=A[1+(|f|/f_k)^\alpha],
\label{eq:noise_pdf}
\end{equation}
where $f_k$ is the knee frequency. We choose $f_k=0.1$~Hz, $\alpha=-2$
and an amplitude $A$ corresponding to the expected white noise level
of each of the 145~GHz PSB receivers; The variation of $A$ between
different receivers is within a factor of two (de Bernardis, priv.\ 
comm.).  The minimum frequency in the TOD is set by the inverse of the
mission life-time, while the nominal sampling frequency of the
experiment (60~Hz) determines the Nyquist frequency. For the sake of
simplicity we assume that no cross-polarisation contamination is
present in the simulations hereafter described. However, we have
performed several tests by varying $\chi_\mathrm{pol}$ up to a $20\%$
level; they all concordantly show that the correct solution is found,
with a small increase in the iteration count and a slightly higher
residual noise level in the final maps. We fix the output map
resolution to $3\farcm 7$ (NSIDE=1024) to properly sample the 145~GHz
beams.

\begin{figure}
  \resizebox{\hsize}{!}{\includegraphics[angle=0]{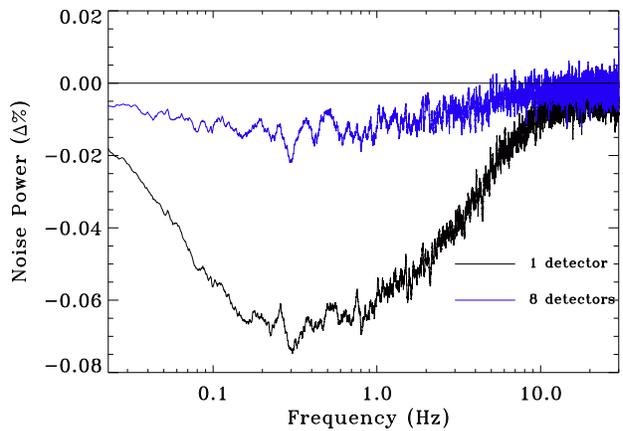}}%
  \caption{Shown, as a function of frequency, is the percentual ratio 
    between the noise power spectrum of a single detector
    estimated from the $S+N$ data (by means of the iterative
    procedure described in the text) and the unbiased estimate of
      the same noise power spectrum obtained from the noise-only
      timestream.  The black (lower) curve is for a single detector
    case, while the blue (upper) curve is obtained with a GLS map
      from all eight detectors. Note the strong suppression of the
    residual bias in this latter case.}\label{fig:noisebias}
\end{figure}

As a first step we perform noise estimation. We follow the iterative
scheme explained above, and we compare the quality of the noise
spectrum recovered when using the single PSB (temperature only) maps
as the signal baseline, as opposed to using the full (global)
temperature map. Not surprisingly, in the latter case, the residual
noise bias is substantially minimised, especially at intermediate
frequencies (see Fig.~(\ref{fig:noisebias})) when compared to the
  unbiased (by definition) estimate of the power spectrum obtained
  from the noise-only TOD used in the S+N simulation.

\begin{figure*}
  \centering
  \includegraphics[width=5.66cm]{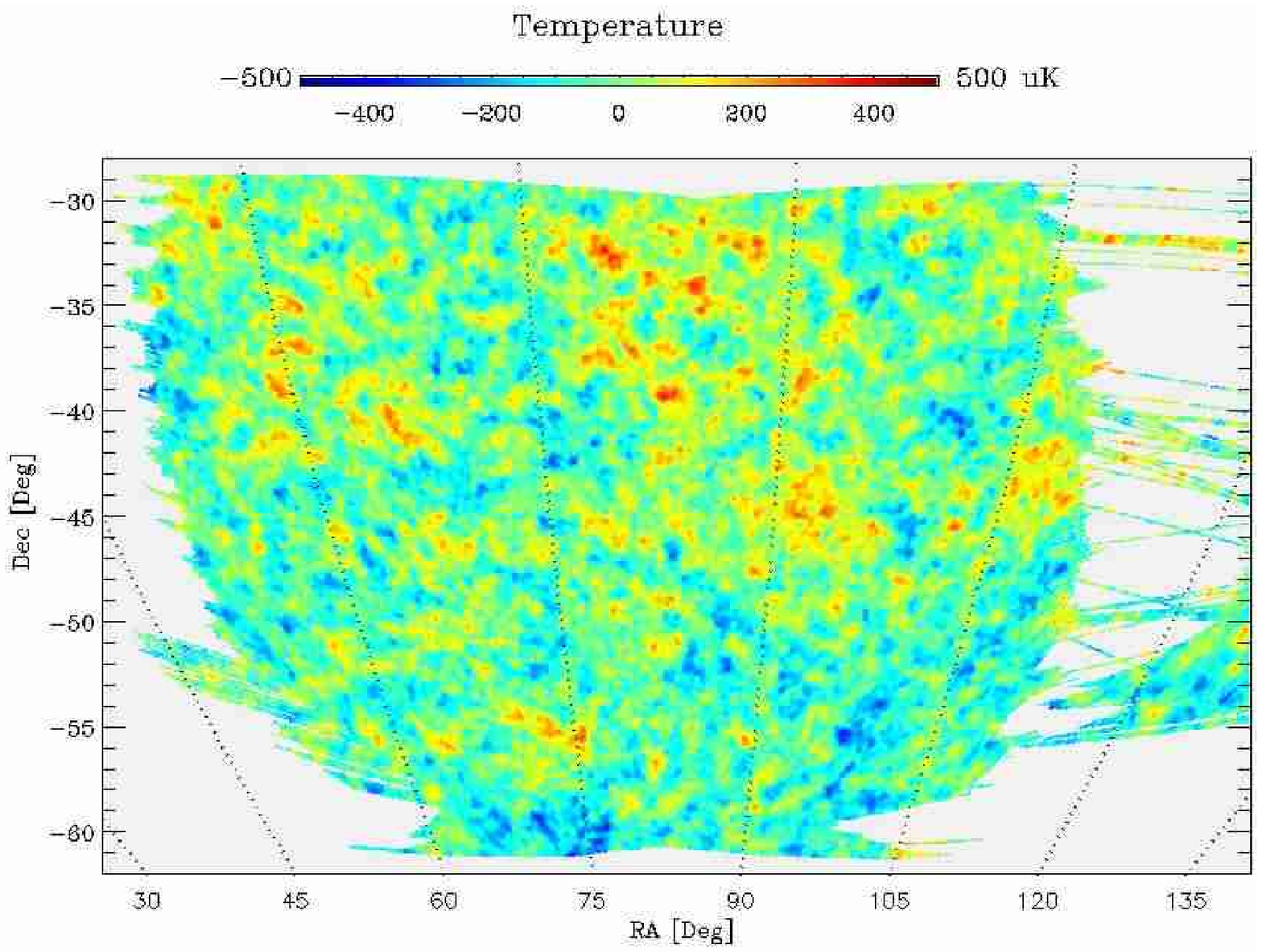}%
  \includegraphics[width=5.66cm]{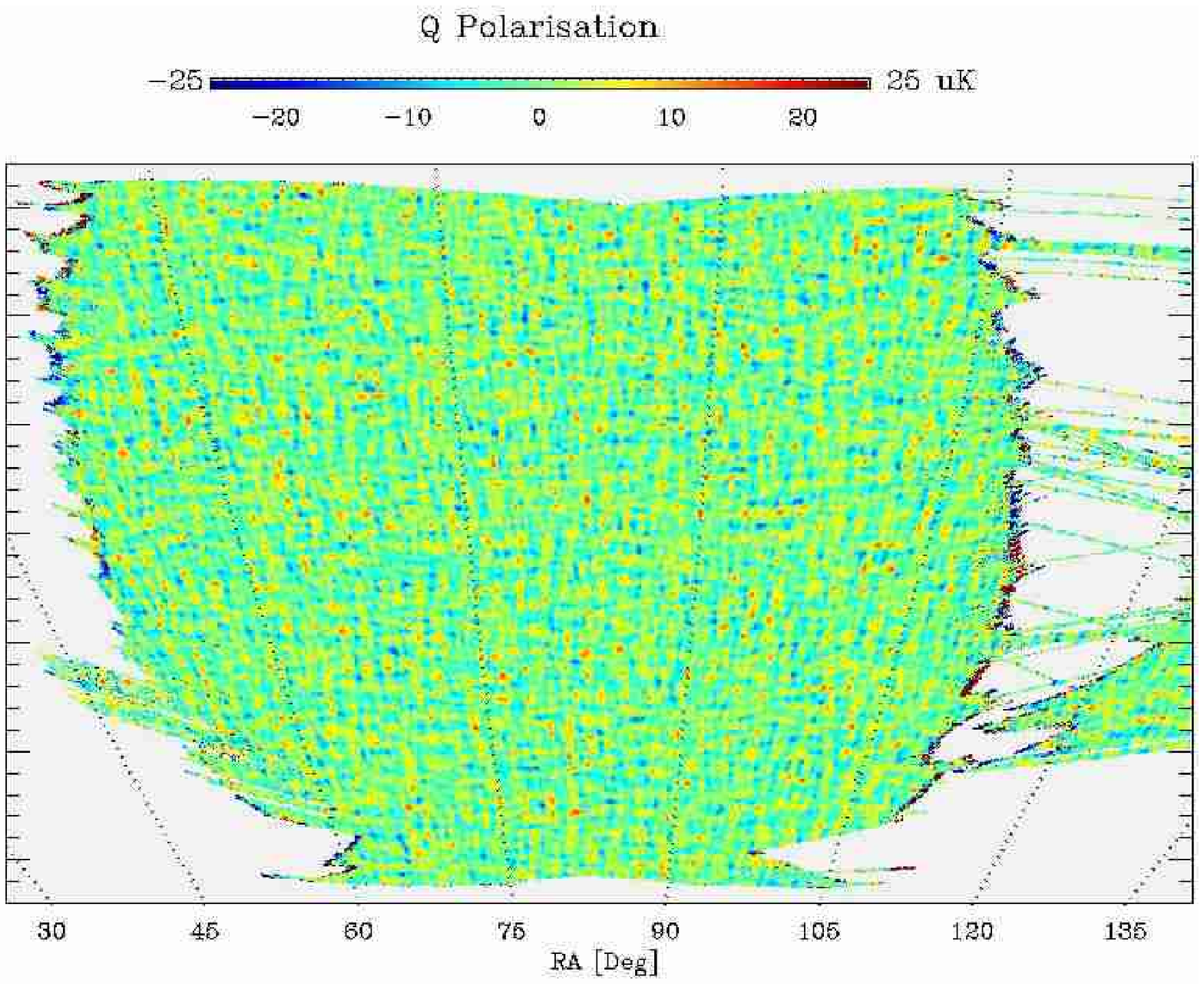}%
  \includegraphics[width=5.66cm]{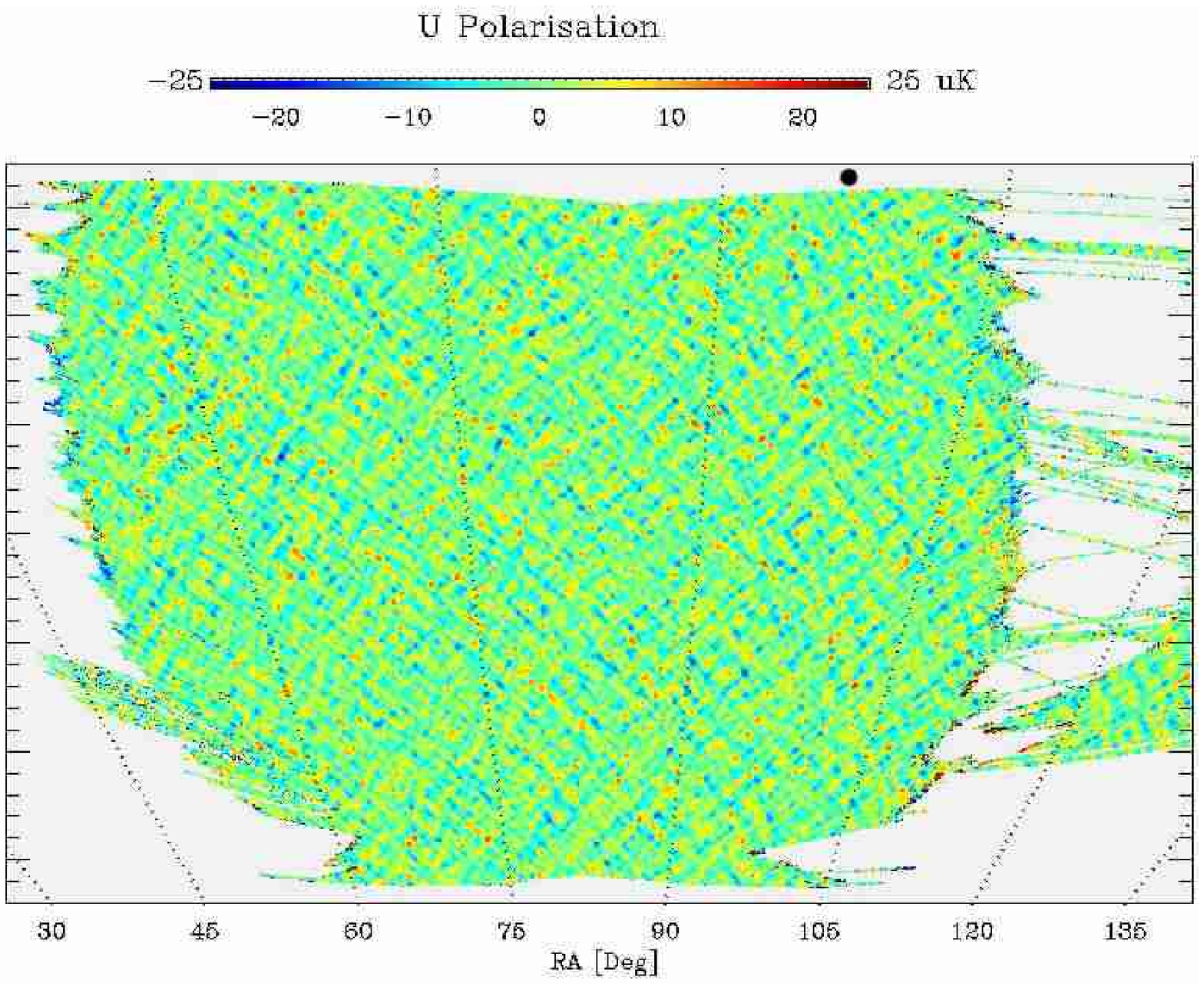}
  \includegraphics[width=5.66cm]{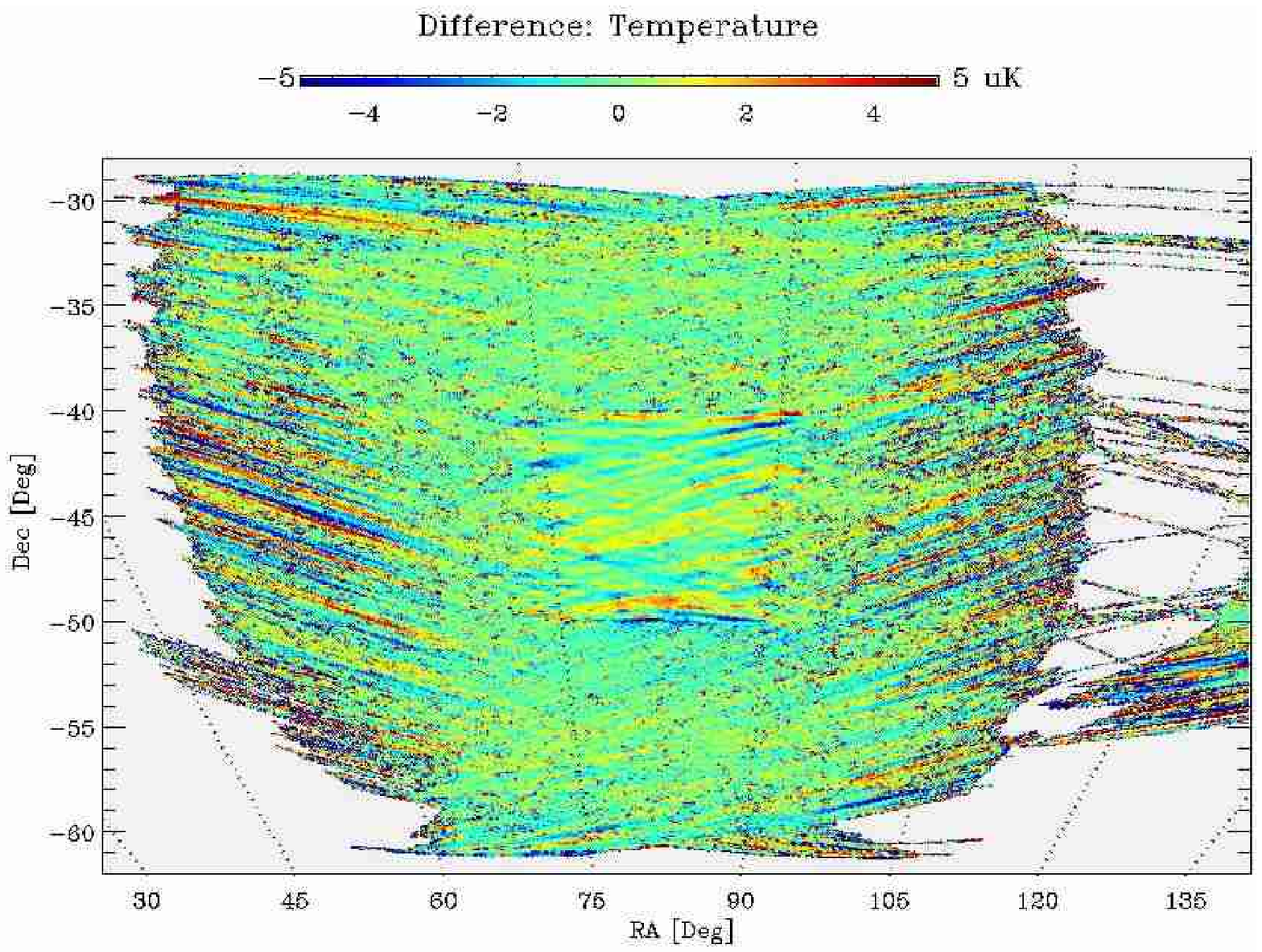}%
  \includegraphics[width=5.66cm]{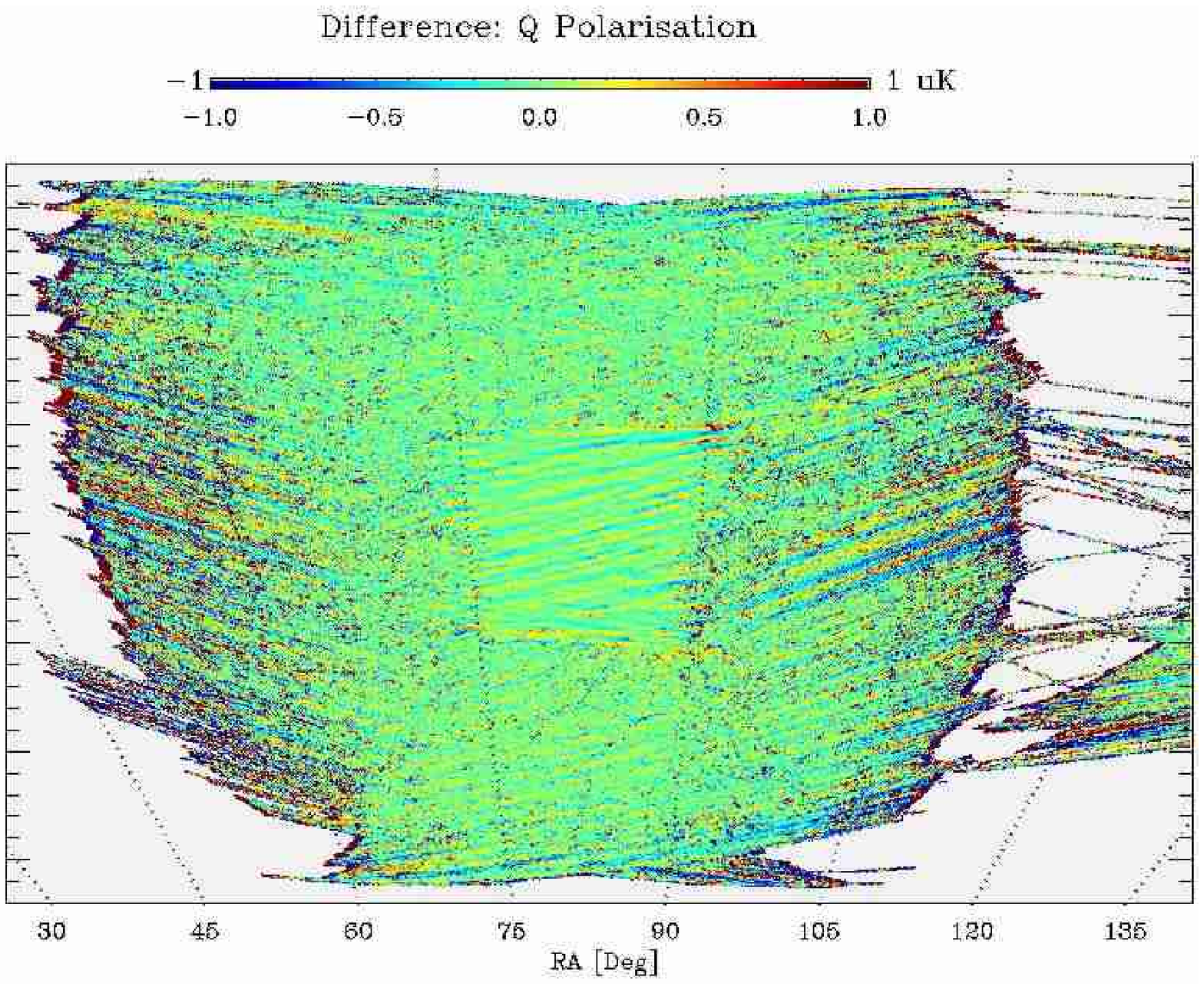}%
  \includegraphics[width=5.66cm]{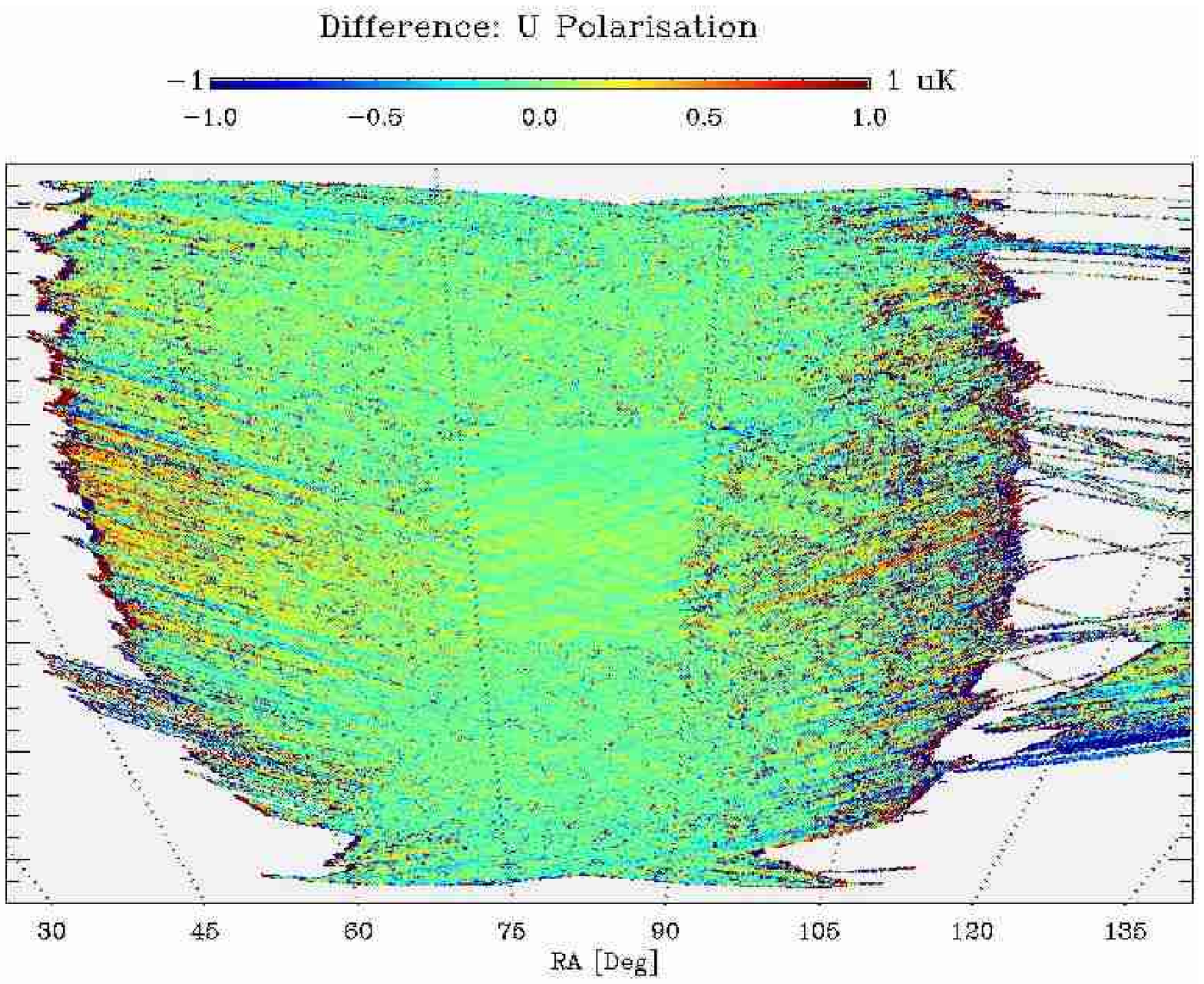}
   \caption{Top row: I,Q and U maps obtained by running ROMA on
     eight ``signal-only'' simulated timelines (one for each B2K
     bolometer).  The display area is restricted to the ``shallow''
     region.  Bottom row: difference maps, computed by subtracting the
     above maps from the simulated sky that was input to the
     simulations. In doing so, the original maps have been degraded to
     NSIDE=1024. Note the change in the colour scale.}
   \label{fig:signal_igls}
\end{figure*}

In Fig.~(\ref{fig:signal_igls}) we display our results for a
signal-only B2K timestream.  Despite the highly realistic simulation
of a complex experiment, the signal maps are recovered to high
precision, better than 1\% for most pixels in the I map, and better
than 10\% for the Q and U maps. Note that the latter result is
consistent, because the standard deviation of the CMB signal in the
polarisation maps is roughly one order of magnitude lower than the one
in the temperature map.  Also note that in the deep region, the high
number of hits per pixel allows to recover the polarisation pattern to
much greater precision.

\begin{figure*}
  \centering
  \includegraphics[width=5.66cm]{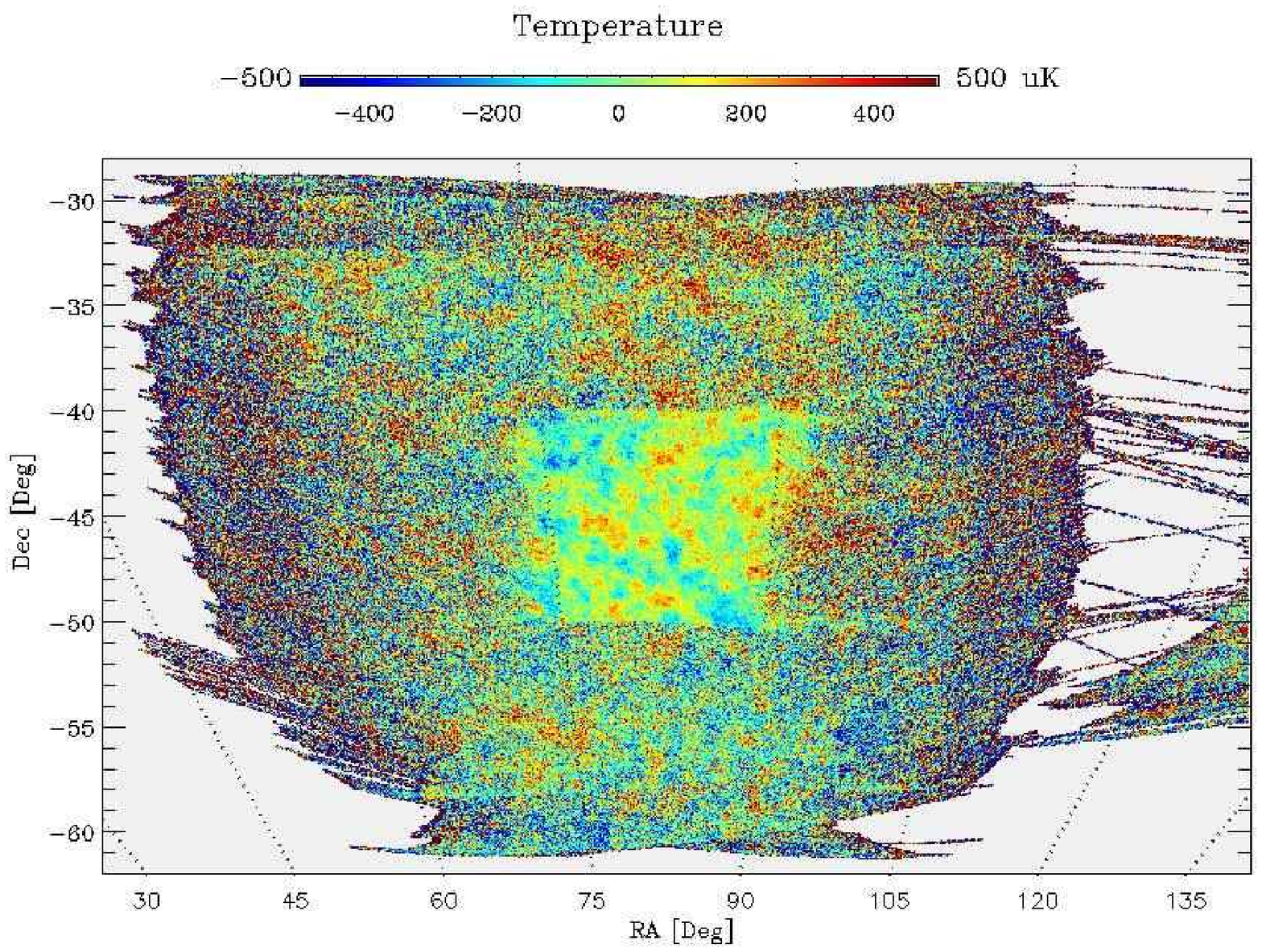}%
  \includegraphics[width=5.66cm]{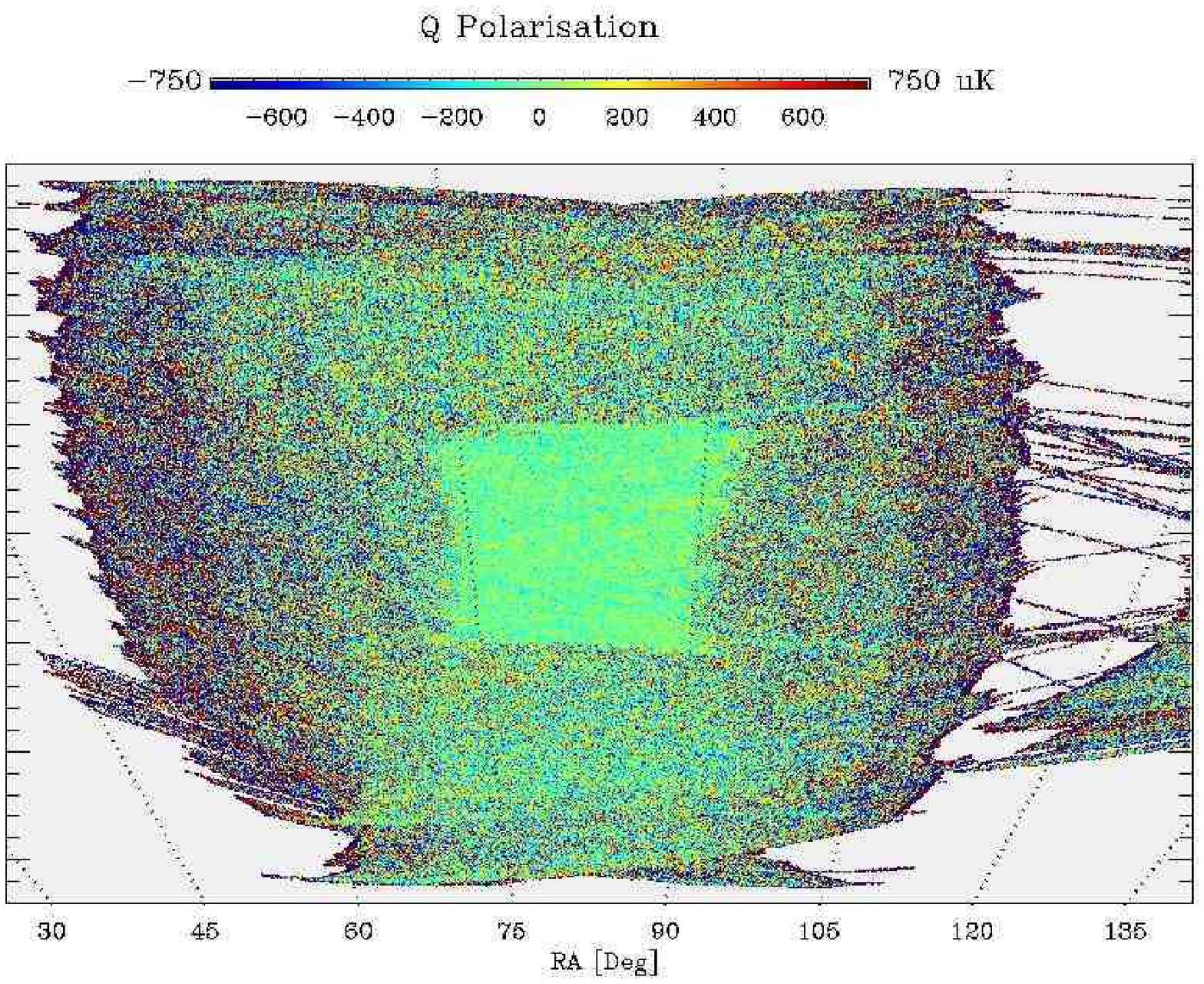}%
  \includegraphics[width=5.66cm]{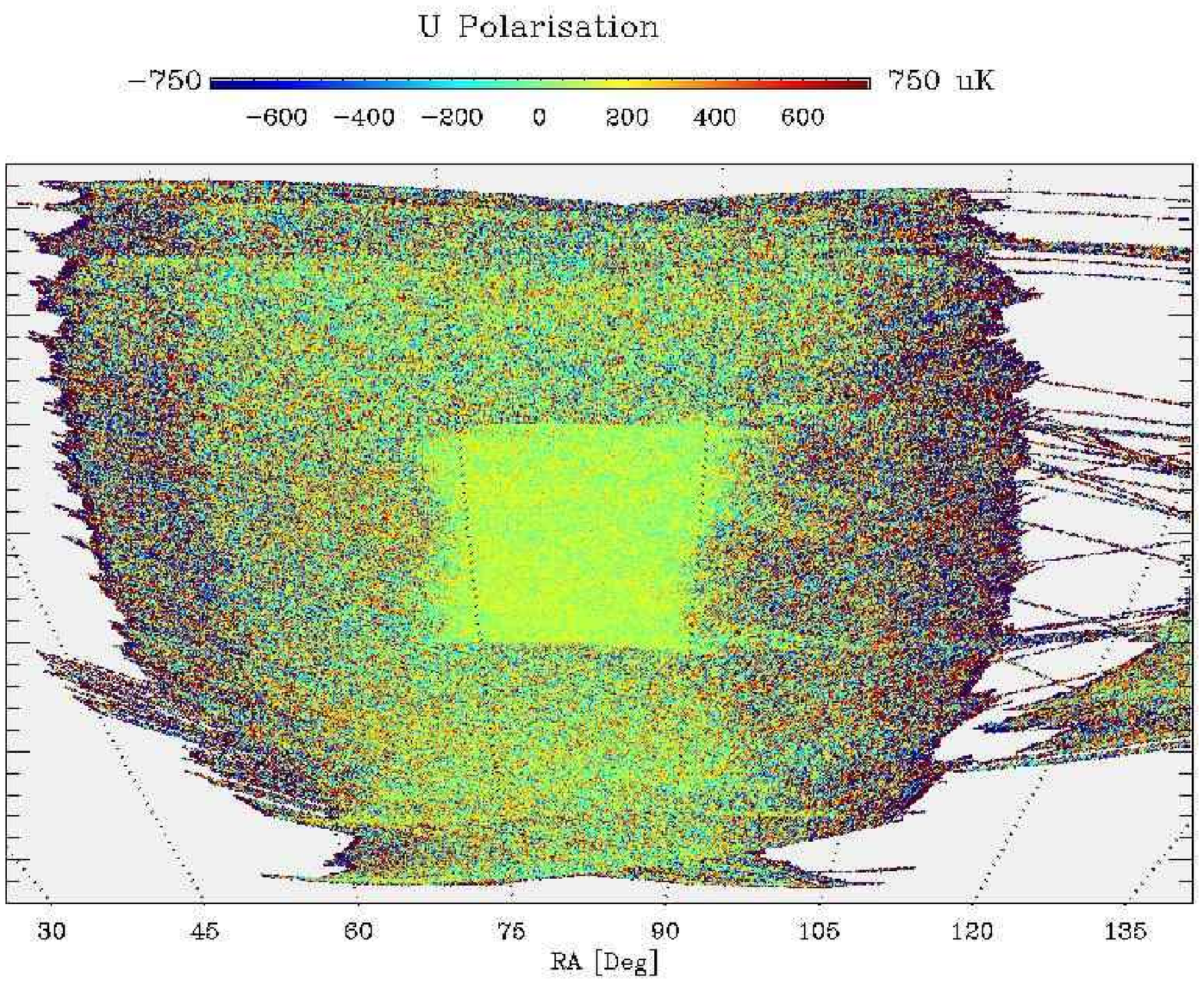}
  \caption{From left to right, I, Q and U output ROMA maps on eight
    ``signal plus noise'' simulated timelines. The assumptions for the
    noise properties are explained in the text.}  \label{fig:sn_igls}
\end{figure*}

The ``signal plus noise'' (S+N) maps are displayed in
Fig.~(\ref{fig:sn_igls}). We have accurately tested that the code
preserves the linearity of the GLS solution it implements: that is,
running ROMA on a S+N timestream is equivalent to summing the output
of two separate S and N runs. Strictly speaking the latter statement
is true only if the code is allowed to iterate until the final
solution is reached. We found in practice that the linearity is very
well preserved after a reasonable number of iteration ($\sim 100$) are
completed.

\begin{figure*}
  \centering
  \includegraphics[width=8.5cm]{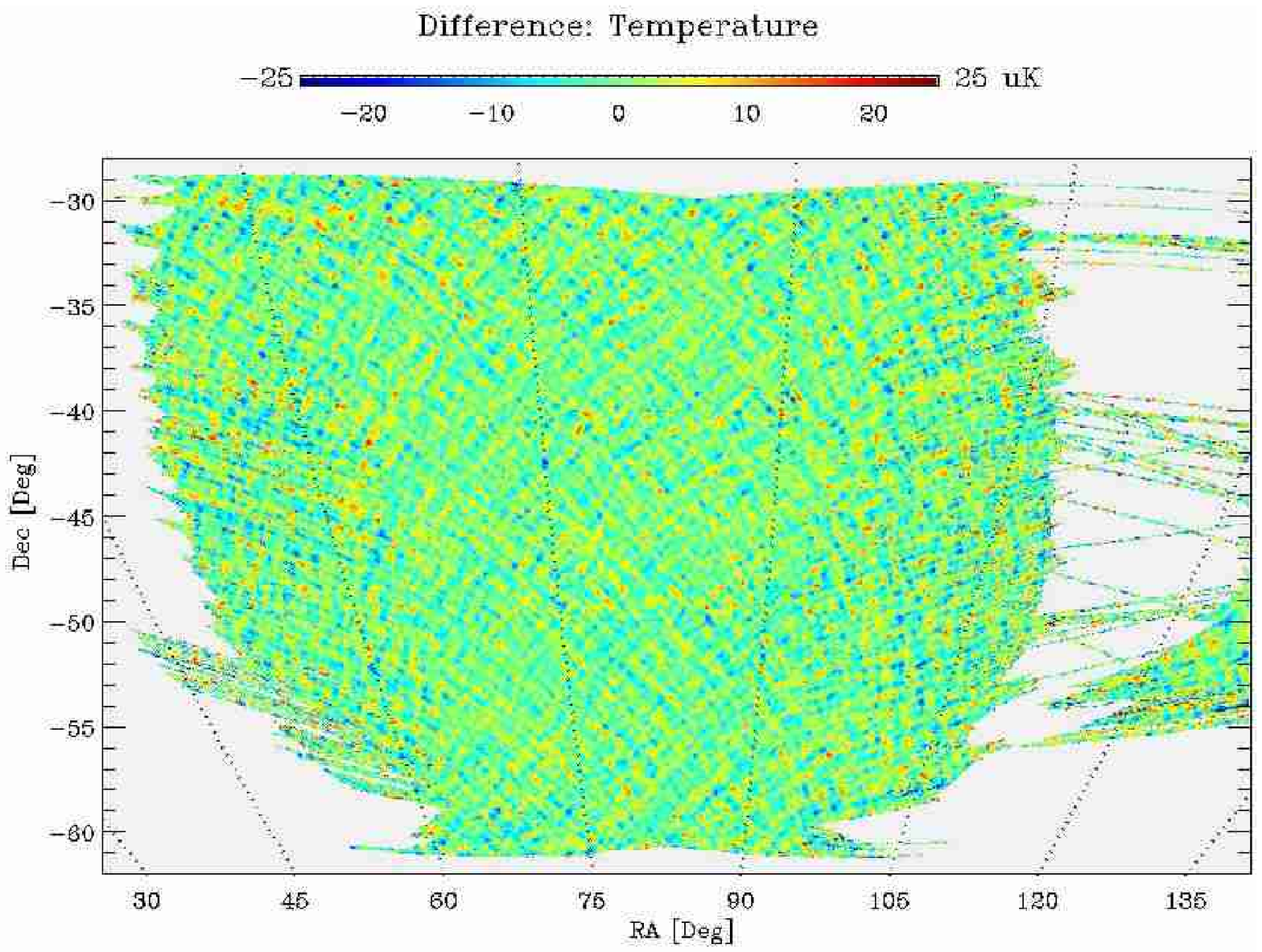}%
  \includegraphics[width=8.5cm]{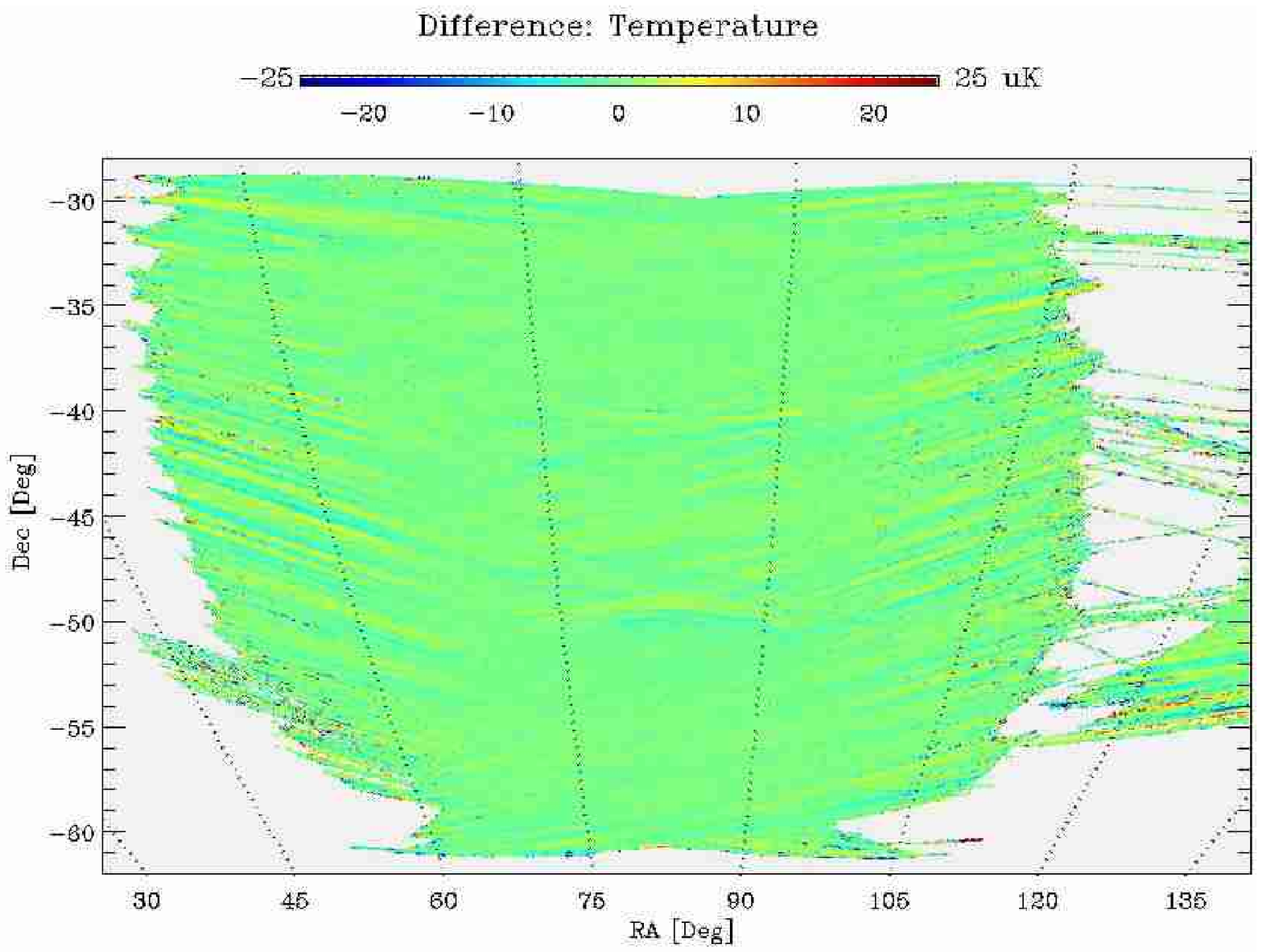}
  \caption{Left panel: Difference between the input, simulated sky and the
    map output by ROMA when processing a single detector,
    ``signal-only'' (\ie\ noiseless) timeline. Both maps (input and
    output) are intensity maps; the input map was degraded to
    NSIDE=1024 before subtraction.  Not surprisingly, the difference
    is dominated by the polarisation pattern, that is present in the
    timeline but cannot be properly reduced when solving for a single
    detector map (that is, no polarisation map can be found). Right
    panel: same as left, but here the $I$ map is estimated jointly
    from all eight detectors ($Q$ and $U$ maps are also computed but
    are not shown here). This figure is \emph{de facto} the one
    displayed in left panel, bottom row of
    Fig.~(\ref{fig:signal_igls}), but on a different
    scale.}\label{fig:signal_T_igls}
\end{figure*}

The importance of having a joint IQU solver is stressed by
Fig.~(\ref{fig:signal_T_igls}).  Here we present, in the same fashion
of the bottom row of Fig.~(\ref{fig:signal_igls}) above, the
difference (input minus output) maps for a signal only case. However
(see also the figure's caption), we show here the residuals obtained
when processing a single channel map, for which no polarisation
solution can be found. As one would expect, this residual is dominated
by a polarisation-like pattern, which the map-making code is unable to
distinguish from the temperature signal.

\begin{figure}
  \resizebox{\hsize}{!}{\includegraphics[angle=0]{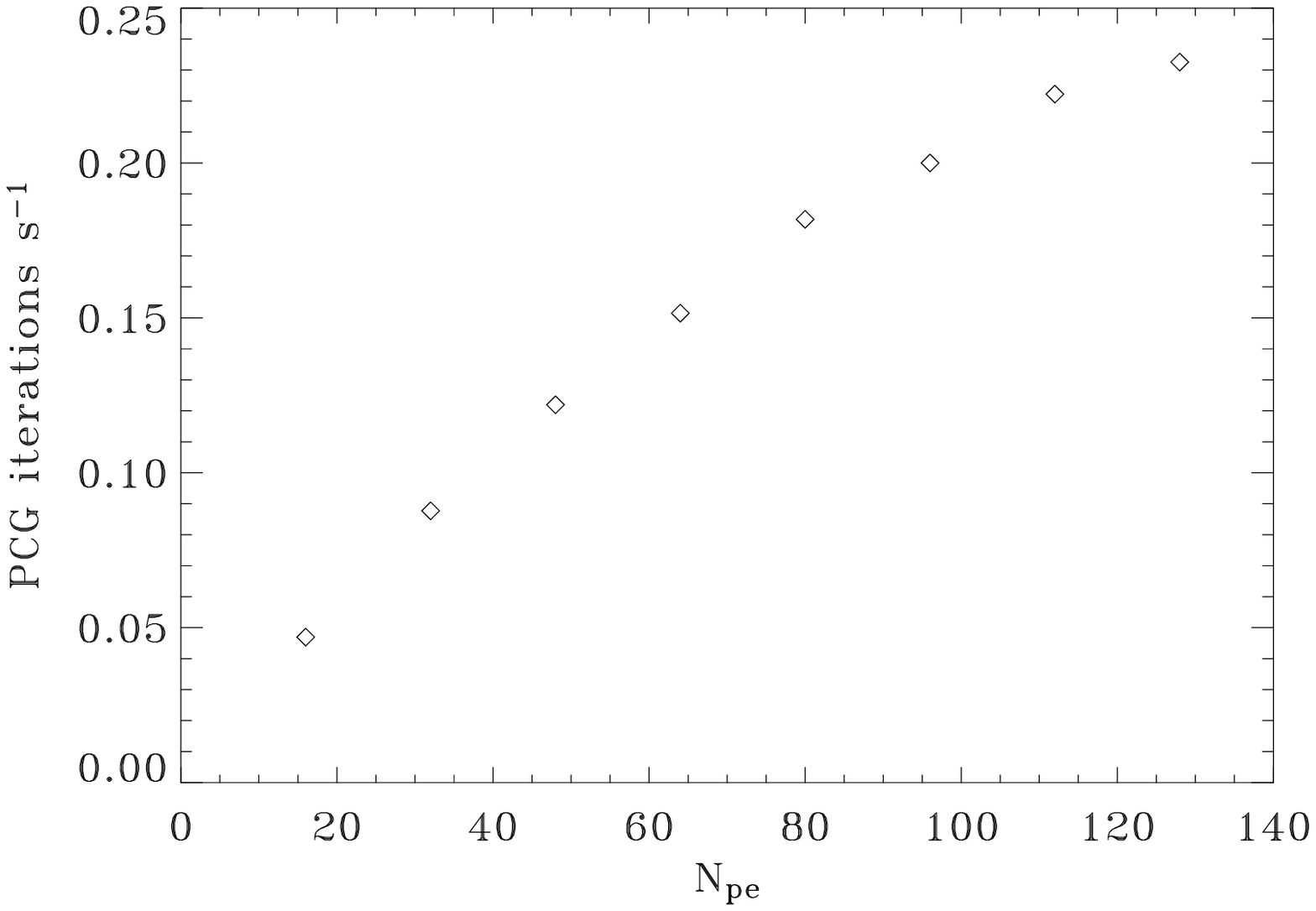}}
  \caption{Shown is inverse time for a single PCG iteration as a function 
    of the number of processor elements ($N_{pe}$) used by ROMA, for
    the full B2K dataset (eight timelines, consisting of $5\times
    10^7$ samples each). The machine used is an IBM Power3 RS/6000.
    ROMA scales optimally until $N_{pe} \sim 80$.}
\label{fig:scaling}
\end{figure}

Most of the computational effort required by ROMA is claimed by the
Fourier transforms, needed to perform repeated convolutions with the
$\mathbf{N}$ matrix. An efficient FFT library must therefore be used.
Our choice falls on the publicly available FFTW3 library
(\cite{fftw}), which claims, in our case, about 80\% of the total
computing time. Use of the FFT guarantees that the code scales
linearly (per PCG iteration) with the number of timeline samples, \ie\ 
with the dataset size.  Strictly speaking, the FFT scales log-linearly
with time samples. However, when performing convolutions, we only take
the non zero band of $\mathbf{N}$, a tunable but constant factor (see
\cite{natoli}). For the B2K test case under consideration, we find
that retaining a noise bandwidth of size $\sim 10^5$ is optimal.  We
find that about $10^2$ PCG iterations are needed to reach convergence
to better than $10^{-5}$ precision in the cases under consideration.
This can be achieved, for instance, in about 5 minutes on a 128
processor job of an IBM SP3, for the full (8 PSB, $4 \times 10^8$
total samples) dataset (see also Fig.~(\ref{fig:scaling})).

\section{Summary and Conclusions}

We have presented our state of the art map-making code to jointly
reduce multichannel CMB anisotropy and polarisation data. ROMA is a
parallel (MPI) implementation of the GLS approach to map-making,
brought to solution by using a PCG iterative method. The only
assumptions are that the optical beam is purely scalar and
axisymmetric, and that the timeline noise is (at least piecewise)
stationary and uncorrelated across different detectors. For the rest,
great care has been taken in tackling real world issues, including
cross-polarisation, multi-detector noise estimation and the problem of
missed data. As a test case we have reduced with ROMA eight PSB
(145~GHz) timelines of highly realistic simulated B2K data. We show
that the IQU maps can be recovered with great precision in the
signal-only case, while attaining the usual GLS (``optimal'') noise
suppression in the noisy case. We stress that, to our knowledge, this
is the first joint temperature and polarisation map-making code
demonstrated to work on a realistic dataset. The code scales linearly
with the dataset size, while its parallel behaviour is quasi-optimal.
It thus represent a viable option to reduce present and forthcoming
large datasets, including \Planck.

\begin{acknowledgements} 
  This research used resources of the National Energy Research
  Scientific Computing Center, which is supported by the Office of
  Science of the U.S. Department of Energy under Contract No.
  DE-AC03-76SF00098. We thank P. de Bernardis and the whole \boom\ 
  collaboration for having provided us with the B2K scan and
  instrumental performances.  We acknowledge use of the HEALPix
  package (\cite{healpix,healpixnew}) and of the FFTW library
  (\cite{fftw}). We thank the CASPUR (Rome-ITALY) computational
  facilities for computing time and technical support. The authors
  wish to thank the \Planck\ CTP working group, and in particular
  C.~M.~Cantalupo and J.~D.~Borril for stimulating discussions.
\end{acknowledgements}

\end{document}